\documentclass[aps, prl, floatfix, %12pt,%
    a4paper,nofootinbib,twocolumn,%
]{revtex4} %% change to aps template
\usepackage[colorlinks,linkcolor=blue,citecolor=blue,urlcolor=blue ]{hyperref}
\usepackage{acronym}

\usepackage{multirow}
\usepackage{graphicx}
\graphicspath{{figs/}}
\usepackage{amsmath,amssymb}
\usepackage{braket}

\usepackage{ulem}
\usepackage{color}

\newcommand{\TRC}{MOE Key Laboratory of TianQin Mission, %
    TianQin Research Center for Gravitational Physics $\&$ School of Physics and Astronomy, %
    Frontiers Science Center for TianQin, %
    Gravitational Wave Research Center of CNSA, %
    Sun Yat-sen University (Zhuhai Campus), %
Zhuhai 519082, China}

\allowdisplaybreaks

\acrodef{gw}[GW]{Gravitational wave}
\acrodef{sbh}[sBH]{stellar mass Black Hole}
\acrodef{smbh}[SMBH]{Supermass Black Hole}
\acrodef{bh}[BH]{black hole}
\acrodef{agn}[AGN]{active galactic nucleus}
\acrodefplural{agn}{active galactic nuclei}
\acrodef{bbh}[sBBH]{stellar mass binary black hole}
\acrodef{bbhs}[sBBHs]{stellar mass binary black holes}
\acrodef{com}[CoM]{center of mass}
\acrodef{df}[DF]{dynamical friction}
\acrodef{mbhbs}[MBHBs]{massive black hole binaries}
\acrodef{mbhb}[MBHB]{massive black hole binaries}
\acrodef{emris}[EMRIs]{extreme-mass-ratio inspirals}
\acrodef{emri}[EMRI]{extreme-mass-ratio inspiral}
\acrodef{imri}[IMRI]{intermediate-mass-ratio inspiral}
\acrodef{imris}[IMRIs]{intermediate-mass-ratio inspirals}
\acrodef{em}[EM]{electromagnetic} 
\acrodef{rm}[RM]{reverberation mapping} 
\acrodef{alma}[ALMA]{Atacama Large Millimeter/submillimeter Array} 
\acrodef{eht}[EHT]{Event Horizon Telescope} 
\acrodef{vlbi}[VLBI]{very long baseline interferometry} 
\acrodef{ztf}[ZTF]{Zwicky Transient Facility} 
\acrodef{xmris}[XMRIs]{extremely large mass-ratio inspirals} 
\acrodef{gbs}[GBs]{Galactic double white dwarf binaries}
\acrodef{gb}[GB]{Galactic double white dwarf binary}
\acrodef{lvk}[LVK]{LIGO-Virgo-KAGRA}
\acrodef{sgwb}[SGWB]{stochastic gravitational-wave background}
\acrodef{smbh}[SMBH]{supermassive black hole}
\acrodef{psd}[PSD]{power spectral density}
\acrodef{snr}[SNR]{signal to noise ratio}
\acrodef{gr}[GR]{general relativity}
\acrodef{fim}[FIM]{Fisher information matrix}
\acrodef{tdi}[TDI]{time delay interferometry}
\acrodef{mhz}[mHz]{milli-Hertz}
\acrodef{mcmc}[MCMC]{Markov chain Monte Carlo}
\acrodef{gp}[GP]{Gaussian process}
\begin{document}

%%========================================================
\title{Probing AGN Disks Density Profiles through Gravitational Wave Observations}
\author{Xiangyu Lyu}

\author{En-Kun Li}
\author{Changfu Shi}
\thanks{Corresponding author: \href{shichf6@mail.sysu.edu.cn}{shichf6@mail.sysu.edu.cn}}
\author{Yi-Ming Hu}
\thanks{Corresponding author: \href{mailto:huyiming@mail.sysu.edu.cn}{huyiming@mail.sysu.edu.cn}}
%\email{corresponding author: huyiming@mail.sysu.edu.cn}

\affiliation{\TRC}

%%======================================================
\begin{abstract}
   Massive black holes surrounded by a gaseous disk have been a prevailing model to explain a wide spectrum of astrophysical phenomena related to \acp{agn}. However, direct and precise measurements of the disk density profiles remain elusive for current telescopes.
   In this work, we demonstrate that it is possible to pinpoint the gas density if an inspiralling stellar mass binary black hole is embedded in the \ac{agn} disk. 
%   A 3-month observation period is sufficient to reach the high precision of density $\delta \rho \sim 10^{-11} \rm g/cm^3$, which is much smaller than the actual density variation $\Delta\rho$ over that period ($\delta\rho \ll \Delta\rho$).
   Furthermore, if the barycenter of the pair follows an eccentric orbit around an AGN, then space-borne gravitational wave detectors can measure the density of the surrounding disk with multi-year observations by tracking the gravitational wave evolution. 
   The error between the inferred density profile and the injected truth can be constrained to below $2\times10^{-11}\rm g/cm^3$.
   Our work opens up an exciting new channel to investigate the very center of galaxies, where disk gas density distributions $\rho(r)$ can be recovered by analyzing time-dependent environmental imprints in gravitational waveforms. 

\end{abstract}



%\tableofcontents

%%@@@@@@@@@@@@@@@@@@@@@@@@@@@@@@
\maketitle

%%======================================================
%%@@@@@@@@@@@@@@@@@@@@@@@@@@@@@@@@@@
%% Introduction
%%@@@@@@@@@@@@@@@@@@@@@@@@@@@@@@@@@@

\textit{Introduction}. -- 
Powered by the gas accretion onto central \acp{smbh}, \ac{agn} system is one of the most luminous objects in the universe~\cite{Rees:1984si}. 
Accretion disks in \acp{agn} serve as the complex engine surrounding central \acp{smbh}, which is vital in researching the coupled evolution of galaxies and central \acp{smbh}~\cite{Fabian:2012xr,Kormendy:2013dxa}.
Numerous efforts have been made to uncover the real structure of the AGN disk, for example, the direct observation from \ac{eht}~\cite{EHT2019ApJ,EHT2022ApJ}, \ac{vlbi}~\cite{baan_h2o_2022} and other ultraviolet/optical continuous spectrum observations~\cite{1978Natur_Shields,Pounds:2003tm,2009ApJ,2021Sci_Burke,2023NatAs_Tang}. 
%And the scientists promoted many theoretical models (e.g., thin disk~\cite{Shakura1973A&A,Sirko:2002ex,Thompson:2005mf}, advection—dominated accretion flow (ADAF) models~\cite{Narayan1994ApJ,Yuan2001MNRAS} and slim disk model~\cite{Abramowicz:1988sp}. %to explain the physics of accretion disks. % are varied in different models. 
However, either the resolution or the detection distances of the above \ac{em} observation methods is inadequate for studying the central regions of \acp{agn}~\cite{EHT2019ApJ,EHT2022ApJ}. %~\cite{Horne2004PASP}. 
Thus, research on accretion disks necessitates new observational approaches. 

With the operation of ground-based \ac{gw} detectors, the recent decade has witnessed a revolutionary breakthrough in exploring the dark universe.
It has been proposed that \ac{bbh} could form and merge in disks~\cite{Morris1993ApJ,Miralda-Escude:2000kqv,Generozov:2018niv,hailey_density_2018}. For example, %AGN channel backed by observation
The \ac{lvk} collaboration has observed \ac{bbh} systems with masses lying in the mass gap~\cite{Belczynski:2017gds,Stevenson:2017tfq,Farmer:2019jed,Mapelli:2019ipt,Marchant:2020haw,Woosley:2021xba,Hendriks:2023yrw}, like GW190521~\cite{LIGOScientific:2020iuh} and GW190426\_190642~\cite{LIGOScientific:2021usb}. The most recent observation GW231123~\cite{LIGOScientific:2025rsn_GW231123} even comprises two component \acp{bh} heavier than 100 $M_\odot$. 
These observations provide strong support for hierarchical mergers, which most likely originate from dynamically active regions, such as the \ac{agn} disks. 
Besides, \citet{Graham:2020gwr} reported that GW190521 ~\cite{LIGOScientific:2020ufj} could be accompanied by an EM counterpart in an AGN system. %could be formed in an AGN system. 
%Their inspiral and merger \ac{gw} signals can be detected by space-borne~\cite{sesana_prospects_2016,liu_science_2020} (e.g. TianQin\cite{luo_tianqin:_2016}/LISA\cite{LISA:2017pwj}/Taiji\cite{Hu:2017mde}) and ground-base \ac{gw} detectors (\ac{lvk})~\cite{LIGOScientific:2021djp,KAGRA:2021duu}.
These \acp{bbh}' \ac{gw} waveforms will encode environmental signatures reflecting disk properties.
For example, the gas density and the SMBH mass~\cite{Cardoso:2019rou,Toubiana:2020drf,DuttaRoy:2025gnu} can be recovered by the corresponding \ac{gw} phase modification by gas accretion~\cite{Caputo:2020irr}, \ac{df} from surrounding gas~\cite{Toubiana:2020vtf}, and binary’s center of mass(CoM) acceleration from the third body~\cite{Bonvin_matter_structure_on_GW,Inayoshi:2017hgw,Tamanini:2016zlh,Yang:2024tje}. 
In conclusion, GW observation is an essential tool for directly measuring the environmental properties. %\acp{gw} penetrate optically thick regions without attenuation, enabling direct measurement of disk characteristics.
However, until now, \ac{lvk} haven't found robust evidence supporting the environmental effects in the detected \ac{gw} signals. %having been detected, no robust evidence supporting the environmental effects in the \ac{gw} signals. 
This absence indicates that the most extreme ambient gas densities $(\ge 10^7\rm g/cm^3)$ have been ruled out for these systems~\cite{CanevaSantoro:2023aol,FedrowPhysRevLett}.

The space-borne \ac{gw} detectors can observe the early inspiral of the \ac{bbh} months to years before these systems enter the high-frequency band ($\ge 10~\mathrm{Hz}$)~\cite{sesana_prospects_2016,liu_science_2020}. 
They can also measure the parameters of the binaries to very high precision:
chirp mass $\Delta M_c/M_c \sim 10^{-7}$, sky localization $\sim 1~\mathrm{deg}^2$, and coalescence time errors $\sim 1~\mathrm{s}$~\cite{sesana_prospects_2016,Sesana:2017vsj,liu_science_2020,Liu:2021yoy,marsat_exploring_2021,Toubiana:2022vpp,buscicchio_bayesian_2021,Lyu:2023ctt}, which facilitates host galaxy identification and enables early warnings for \ac{em} follow-up observations~\cite{sesana_prospects_2016,liu_science_2020,Zhu:2021bpp}.
Besides, parameter of disks like gas density $\rho$ can be well constrained that $\delta \rho \sim 10^{-10} \rm g/cm^3 - 10^{-12} g/cm^3 $~\cite{Toubiana:2020drf,Caputo:2020irr,DuttaRoy:2025gnu}. 
We remark that previous studies assume a stable environment, and only local properties of the disk can be recovered.

By adopting the thin disk model~\cite{Sirko:2002ex,Thompson:2005mf}, the gas density $\rho$ decreases with larger radial distance $r$ to \ac{agn} center $\rho\propto r^{-3}$. 
This means that if the barycenter of \ac{bbh} moves around the \ac{agn} center on an eccentric orbit, these systems may experience the variation of the density $\Delta \rho$ .
The X-ray observation of the quasiperiodic eruptions (QPEs) often reports a “long-short” pattern for the recurrence time, and some suspect that compact objects that follow an eccentric orbit around the massive black holes might just explain this~\cite{Jiang_2025,Linial:2023nqs}. 
%Under this circumstance, \acp{bbh} \ac{gw} could encode information about the disk profile.

In this letter, we demonstrate that space-borne detectors, such as TianQin, can achieve high-precision constraints on the density profile of the AGN accretion disk through \ac{bbh} GW observations. 
We show that, under the assumption that the barycenter of the \ac{bbh} follows an eccentric orbit, the parameter estimation results for the environment density are precise enough to distinguish the inherent gas density variation over three-month observational intervals, and the profile of the disk can be determined with sequential observations. In this study, we employ a natural unit system with $c=G=1$.

%%@@@@@@@@@@@@@@@@@@@@@@@@@@@@@@@@@@
%% sBBHs in AGN disks
%%@@@@@@@@@@@@@@@@@@@@@@@@@@@@@@@@@@

\textit{Capability of constraining environmental effect by \ac{bbh} systems}. --

\acp{bbh} will experience \ac{df} effect throughout their evolution if embedded in a dense environment ({\it e.g.}, accretion disk or dark matter halo), while the environment could imprint detectable signatures on \ac{gw} signals. 
This effect exerts a drag force on component BH, introducing a $-5.5$ post-Newtonian order phase correction to the waveform~\cite{Enrico_accretion_inference_EMRI,Barausse_environmental_astrophysics,Caputo:2020irr,Kocsis_accretion_DF}: 
\begin{equation}
    \tilde{\phi}_{\rm DF} \simeq -\rho\frac{25\pi(3\eta-1)M_z^2}{739328\eta^2}\gamma_{\rm DF}[\pi fM_z]^{-16/3},
    \label{equ:DF effect}
\end{equation} %The \ac{df} 
where $\rho$ is the ambient density, $M_z = M_c(1+z)$ is the redshifted chirp mass of \ac{bbh} system, $\eta$ is the systematic mass ratio, $\gamma_{\rm DF} $ is the relative \ac{df} effect parameter~\cite{Toubiana:2020drf}. 

% choice of parameters
In our analysis, we adopted a GW190521-like \ac{bbh} as the fiducial system with the source parameters 
$\boldsymbol{\theta}= \{ D_L, M_c,\eta, \chi_1,\chi_2,\iota, t_c,\phi_c \}$ chosen as: luminosity distance $D_L=1000 \rm Mpc$, chirp mass $M_c= 77.66M_\odot$, systematic mass ratio $\eta=0.23$, aligned spin $(\chi_1=0.80,\chi_2=0.80)$, inclination angle $\iota=\pi/6$, coalescence phase $\phi_c=0$, and coalescence time $t_c = 5\text{yrs}$. 
The \ac{df} effect parameter $\gamma_{\rm DF} =  -247 \log \left[ \frac{f}{c_s / (22 \pi M)} \right] - 39 + 304 \log 20 + 38 \log \frac{3125}{8}$ is related with the gas sound speed $c_s$ and total mass $M$, which was set to be $0.02$ and $155.60M_\odot$.
We neglected the sky location and polarization angle and instead used the sky-averaged TianQin antenna pattern function~\cite{liu_science_2020} and the IMRPhenomPv2~\cite{IMRPhenomPv2} waveform. 
Specially, TianQin has a ``3 month on, 3 month off" observation schedule~\cite{luo_tianqin:_2016}, however, for the purpose of demonstration,
%in order to demonstrate the space-borne \ac{gw} detectors' capability of constraining environmental effect,
we assume a 5yr continuous observation $(T_{\rm orb}=5 \rm yrs)$, and the observation data $\boldsymbol{D}$ compose of 20 segments $[D_1,D_2,\dots,D_{20}]$.  
Based on the fiducial system, we adopted the same environment density $\rho=10^{-10}\rm g/cm^3$ in ~\citet{Toubiana:2020drf} as the injected value in a three-month simulated \ac{gw} signal, and we assume this value to be a constant during this period.

% basics on calculation methods
To assess capability of TianQin in constraining environmental effect, we first simulate a five-year mock observation data $\boldsymbol{D}$, and we do not inject noise to avoid potential bias. 
Under the aforementioned parameter set $\boldsymbol{\theta}$, the \ac{snr} of the fiducial system is $11.8$, which surpasses the detection threshold (\ac{snr}=8) of space-borne \ac{gw} detectors.
We then estimate the posterior distribution $P(\boldsymbol{\theta}|D)$ under the Bayesian inference, with the first three-month observation data $D_1$. In Bayesian inference, the posterior distribution $P(\boldsymbol{\theta}|D)$ is proportional to the product of the likelihood function $\mathcal{L}(\boldsymbol{\theta})$ and prior $P(\boldsymbol{\theta})$. 
The logarithm of likelihood $\mathcal{L}(\boldsymbol{\theta}) $ can be defined as $ \log\mathcal{L}(\boldsymbol{\theta})\propto -\frac{1}{2} \langle D-h(\boldsymbol{\theta})|D-h(\boldsymbol{\theta})\rangle$, where $\langle a(t),b(t)\rangle = 4\Re \left[\int_0^{\infty}{\rm d}f \frac{\tilde{a}(f)\tilde{b}^*(f)}{S_n(f)} \right]$ is the inner product, with the $S_n(f)$ being the one-sided \ac{psd} \cite{liu_science_2020,WangHT:2019ryf}.

% PE methods introduction
For the Bayesian inference, we adopt the \texttt{emcee} implementation~\cite{foreman-mackey_emcee_2013} of the affine-invariant \ac{mcmc} ensemble sampler.
Also use the \ac{fim} method to calculate the precision $\delta \theta$ and therefore estimate the ability of TianQin in constraining the gas density.
With the \ac{fim} matrix $\Gamma_{ij}= \langle \frac{\partial h}{\partial\theta_i}, \frac{\partial h}{\partial\theta_j}\rangle$, 
the covariance matrix can be approximated as $\Sigma = \Gamma^{-1}$~\cite{cutler_gravitational_1994}. 

%%===================================================
\begin{figure}[htbp]
    \centering
    \includegraphics[width=\linewidth]{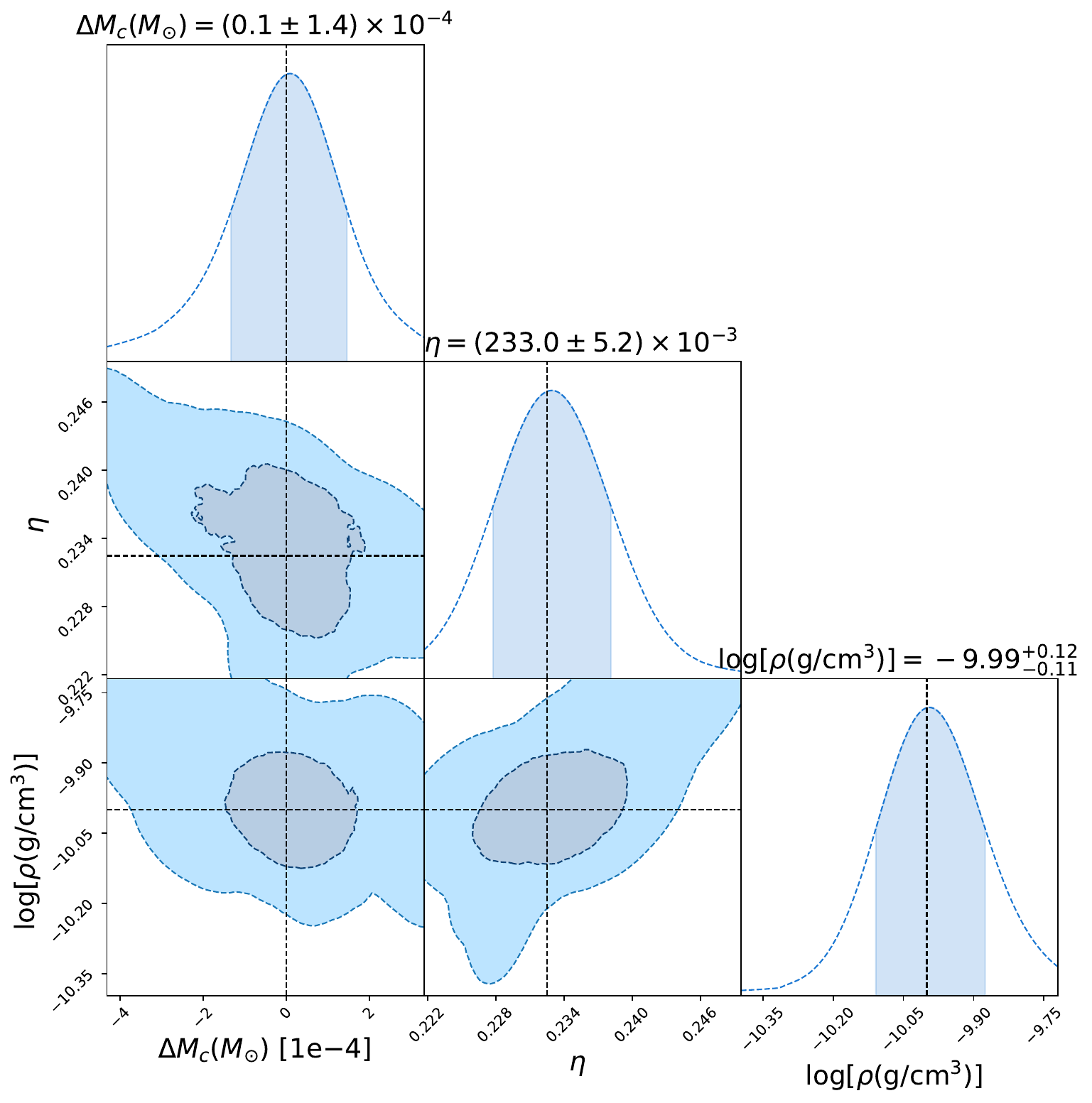}

    \caption{ The posterior distribution of three parameters $\Delta M_c,\eta,\log\rho$ of a GW190521-like system. Black dashed lines represent the injected value, and light blue contour lines are 68\% and 95\% credible intervals (CI). And subplot headers report central estimates with $1\sigma$ uncertainties. 
    }
    \label{fig:3D_posterior}
\end{figure}
%%==================================================

Fig.\ref{fig:3D_posterior} demonstrates the parameter estimation results for $(M_c,\eta,\rho)$ under three-month observation of TianQin, with all other parameters fixed to the injected values, and $(M_c,\eta,\rho)$ are the only free parameters, while all other parameters are fixed to their injected values.
For clearer visualization, we present the relative deviation in chirp mass $\Delta M_c = M_c - M_c^{\rm inject}$, and express the environmental density in logarithmic form $\log[\rho/{\rm (g/cm^3)}]$. 
Our results reveal that $\rho=0~{\rm g/cm^3}$ is outside the $90\%$CI region, highlighting the fact that TianQin could decisively detect the environment effect in the \ac{gw} signal.
Furthermore, with three-month observation, TianQin can constrain the environment density to the precision $\delta\rho\sim 10^{-11}\rm g/cm^3$.
%The value is nearly one order larger than the previous studies~\cite{Cardoso:2019rou,Toubiana:2020cqv,DuttaRoy:2025gnu}, which can be explained by the multi-year observation time adopted in previous studies.

%%@@@@@@@@@@@@@@@@@@@@@@@@@@@@@@@@@@
%% GW observations to AGN
%%@@@@@@@@@@@@@@@@@@@@@@@@@@@@@@@@@@

\textit{Recovering the disk density profile}. -- Throughout five-year operational lifetime of TianQin~\cite{luo_tianqin:_2016}, sequential three-month observational segments could provide independent environmental density estimates $[\rho_1\pm\delta\rho_1,\rho_2\pm\delta\rho_2,\dots,\rho_{20}\pm\delta\rho_{20}]$. 
In this study, we assume an AGN accretion disk as the astrophysical environment for \ac{bbh} systems.
If the barycenter of the \ac{bbh} follows an eccentric orbit, then the changing radial distance leads to changing density. 
The density variation between two segments is quantified as $\Delta\rho_i = |\rho^{(i)} - \rho^{(i-1)}|$, representing the absolute difference of densities at adjacent observation segments.

%{Introduce how to constrain the profile}
In order to calculate the actual value of the $\Delta \rho_i$, we adopt a number of assumptions on the underlying accretion disk.
%To quantify these density variations $\Delta\rho_i$ throughout the whole observation period,
For example, we employ the Sirko-Goodman (SG) thin disk model~\cite{Sirko:2002ex}.
In the SG model, accretion disk could be partitioned into two distinct regions: the inner region is geometrically thin, with high angular frequency and temperature~\cite{Pringle:1981ds}, the Toomre parameter $Q \approx \Omega^2/2\pi \rho \gg 1$~\cite{Toomre1964ApJ}, and the outer region is a self-gravitating disk near marginal stability, where $Q \approx 1$~\cite{Toomre1964ApJ}, with $\Omega = \sqrt{M_{\rm SMBH}/r^3}$ as the angular velocity of the disk.%, $r_s\equiv 2M_{\text{SMBH}}$. 
In the outer region of the disk, the gas density profile $\rho(r)$ follows~\cite{Sirko:2002ex}: $\rho(r) = \Omega^2/2\pi=  M_{\text{SMBH}}/(2\pi r^3)$.
The power-law distribution of gas density in the accretion disk allows the density difference $\Delta\rho_i$ to vary in different observation segments.

The key to recovering the density profile is to associate the density $\rho$ with the corresponding radius $r$.
Following an eccentric orbit around the central \ac{smbh}, the barycenter of the \ac{bbh} to the \ac{smbh} evolves as time $r(t)$, and its exact evolution is determined by the semi-major axis $a$ and the orbit eccentricity $e_{\rm orb}$.
Therefore, by assuming that the density is closely linked to the radius and by monitoring the changing rate of environmental density, it is possible to determine the orbital elements, such as the orbital eccentricity.

For the precision of gas density $\delta\rho_i$ estimation, we approximate with the \ac{fim} method.
While the uncertainty in radius roots in the uncertainty in the timing.
% For the AGN disk, we adopt the parameters from the ZTF19abanrhr \cite{Graham:2020gwr}, where the \ac{smbh} mass $M_{\rm SMBH}$ is $10^8M_{\odot}$ and the radial distance of is $\sim 700r_g,r_g=M_{\rm SMBH}$. And the eccentricity of \ac{bbh} barycenter's orbit $e_{\rm orb}$ was set to be $0.3$. % as this non-zero value is permitted in the environment of the AGN disk.
For the convenience of calculation, throughout our analysis, we assume the gas density to be constant during the $i$-th three-month period $\rho_i$, and there must be a moment $\overline{t_i}$ when $\rho(\overline{t_i})=\rho_i$, so that an effective constant density can explain the actual accumulated phase shift due to the changing gas density.
Since in practice we do not know the exact time, we use the mid-time $t_i^{\rm mid}$ to approximate, and we denote the difference between $\overline{t_i} $ and $t_i^{\rm mid}$ as $\Delta t_i$.
We take $\sigma_t$, the standard deviation of the 20 $\Delta t_i$, to indicate the timing uncertainties.
Combined with the determined orbit, the timing uncertainties can be translated into radius uncertainties.
%And we assume the environment surrounding the \ac{bbh} system follows the power-law distribution and maintain the assumption of the static profile during observation, which enables the unambiguous attribution of detected density variations to orbital motion rather than temporal disk evolution. 
%{Introduce the system bias introduced by the set of the time of ρ_i in each segment}
%Because, over a typical three-month observing period, the intrinsic variation in gas density, $\Delta \rho_i \sim 10^{-8} \rm g/cm^3$ can indeed exceed our detection ability $\delta \rho\sim 10^{-10} \rm g/cm^3$. 
%The process that we designate the midpoint $t_i^{\rm mid}$ of each observation window as the representative time for the gas density $\rho_i$, this fails to capture the dynamic nature of the disk. 

%{Detailed introduce the time bais from our midpoint density assumption}

We combine all the previous models to determine both the \ac{bbh} orbital elements ($a,~e_{\rm orb},~ \phi_0)$, where $\phi_0$ is the initial orbital phase), as well as the disk density profile parameters ($\alpha,~M_{\rm SMBH}$)using the sequential observation of the \ac{bbh} inspiral waveforms $\boldsymbol{\rho}=[\rho_1\pm\delta\rho_1,\rho_2\pm\delta\rho_2,\dots,\rho_{20}\pm\delta\rho_{20}]$ and the corresponding observation time $\boldsymbol{t} = [t_1\pm\sigma_t,t_2\pm\sigma_t,\dots,t_{20}\pm\sigma_t]$.
We assume the estimation of the gas density and the time are statistically independent; therefore we construct the likelihood function $P(\boldsymbol{\rho,t}|\bar{\theta})$ as:
\begin{equation}
    -\frac{1}{2}\log P(\boldsymbol{\rho,t}|\bar{\theta}) = \sum_i \frac{\left(\rho_i-\rho(t,\bar{\theta})   \right)^2}{2\delta\rho_i^2} + \sum_i \frac{\left(t_i-t   \right)^2}{2\sigma_t^2},
    \label{equ:chi square}
\end{equation}
where $\rho(t_i,\bar{\theta})=\rho_0 [r(t_i,a,e_{\rm orb},\alpha,M_{\rm SMBH},\phi_0)/r_0]^{-\alpha}$, $r_0 = 350r_s$~\cite{Graham:2020gwr}, $r_s\equiv 2M_{\text{SMBH}}$ is the Schwarzschild radius of central SMBH.
Time series $t_i,i=1,2,\cdots,20$ are obtained from the \ac{gw} observation segments, and the time-dependent radial distance $r$ can be obtained by $(a,e_{\rm orb},\alpha,\phi_0)$.  
Finally, we can construct the posterior function from the likelihood and prior as 
Thus, we adopted $\chi^2$ as our likelihood function $\mathcal{L}(\bar{\theta})$ to calculate the posterior distribution of $\bar{\theta}=a,e_{\rm orb},M_{\rm SMBH},\phi_0$. $P(\bar{\theta})$ is the prior of the parameters $\bar{\theta}$
\begin{equation}
    P(\bar{\theta}|\boldsymbol{\rho},\boldsymbol{t}) \propto P(\bar{\theta})P(\boldsymbol{\rho}, \boldsymbol{t} |\bar{\theta})
\end{equation}

%%===================================================
\begin{figure}[htbp]
    \centering
    \includegraphics[width=1.0\linewidth]{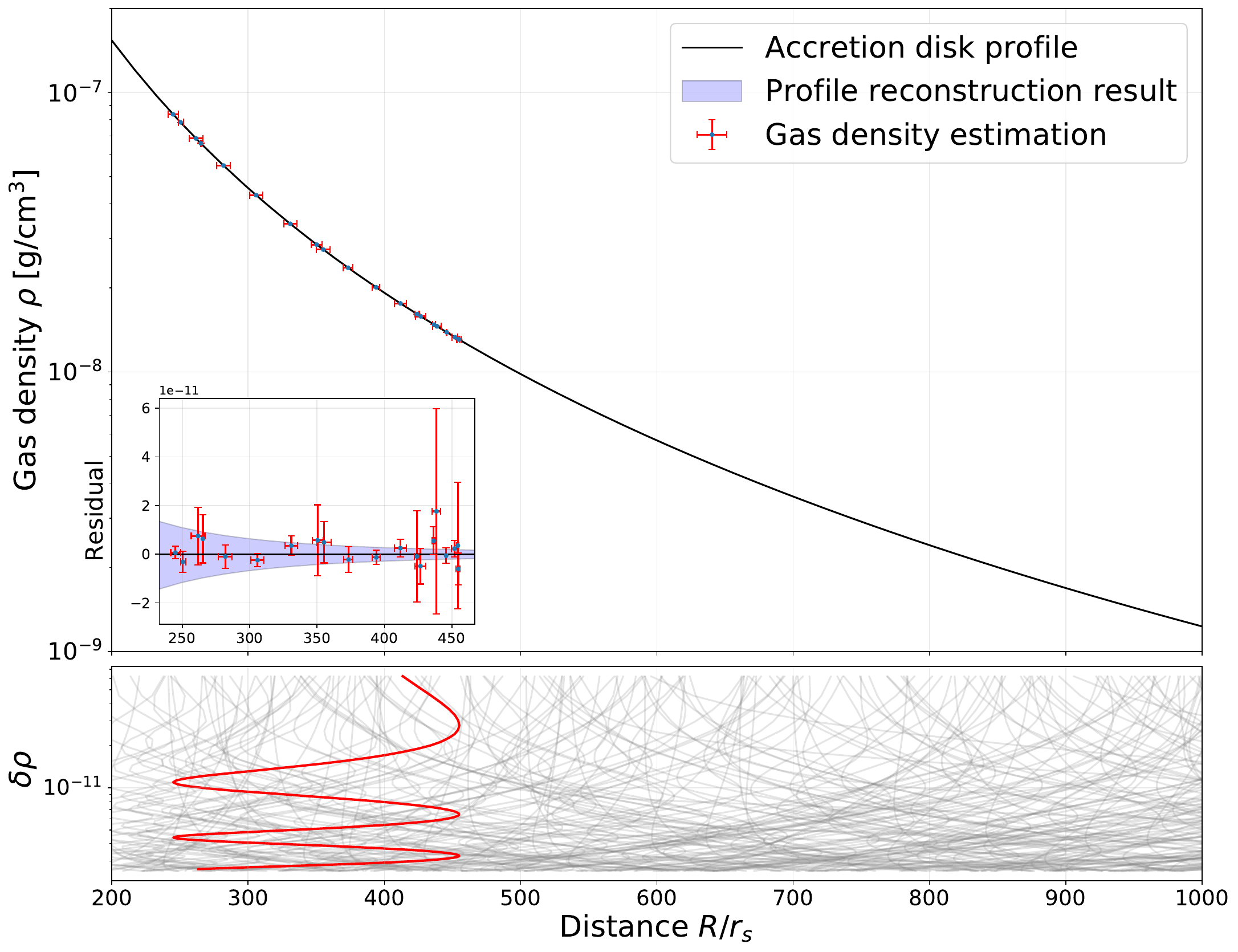}
    \caption{ The upper panel shows the injected (black line, SG model) and recovered (blue shaded region) disk density profile with an example event. Dots and error bars indicate the estimate of gas density and radial distance of the 3-month segments. The uncertainties are so small that we highlight the difference between the injection and the recovery in the zoomed-in panel.
    The lower panel shows the evolution of the gas density precision $\delta\rho$ for a population (grey) and the example (red) GW190521-like event.
    }
    \label{fig:FIM_rho}
\end{figure}
%%==================================================
% upper panel, introduce error bar and zoom-in
In Fig.~\ref{fig:FIM_rho}, we present our reconstruction ability on the density profile of the AGN disk.
We simulate the data $\boldsymbol{\rho},~\boldsymbol{t}$ by adopting a $10^8M_{\odot}$ central \ac{smbh} and a semi-major axis $a\sim 350r_s$.
This combination lead to a period $T = \sqrt{4\pi^2a^3/GM_{\text{SMBH}}}\approx 1.816\rm \;yrs$.
The orbital eccentricity of the \ac{bbh} barycenter $e_{\rm orb}$ is set to $0.3$. 
We then obtain the posterior through \texttt{emcee}.

In the upper panel, we plot the 20 estimates of the environmental densities, with the horizontal axis indicated by the radius during the three-month observation time.
The density is shown together with their uncertainties $\rho_i\pm\delta\rho_i$ with error bars. 
In the horizontal axis, we show the radial distances of the \ac{bbh} system to the central \ac{smbh},
 $r_i \pm \Delta r_i$ in the Schwaichii radius $r_s$ unit.
This is not the direct observables, but we convert the timing information to radius for better visualizations.
According to the posterior, we also draw the uncertainties of the disk density profile with blue shaded regions.
However, the uncertainty is too small, and we provide a zoomed-in subplot within to highlight the precision, using the blue shaded region to indicate the 90\%CI of the residual between the recovered and the disk profile. 
From the plot we can observe that the \ac{agn} disk profile can be constrained with a precision generally better than $2\times10^{-11}\rm g/cm^3$.
In the bottom panel, we show the evolution of estimation precision $\delta\rho$ (assuming a sliding 3-month window around) for a population of events. 
All of them are assuming the same GW190521-like binary parameters, but with varying $(a,e_{\rm orb})$ combinations around the \ac{smbh}. 
The example event adopted in the upper panel is highlighted in red.
We can observe that as the \ac{bbh} continues to spiral in, it decouples from the surrounding gas, so that the uncertainty increases.

%%===================================================
\begin{figure}[htbp]
    \centering
    \includegraphics[width=\linewidth]{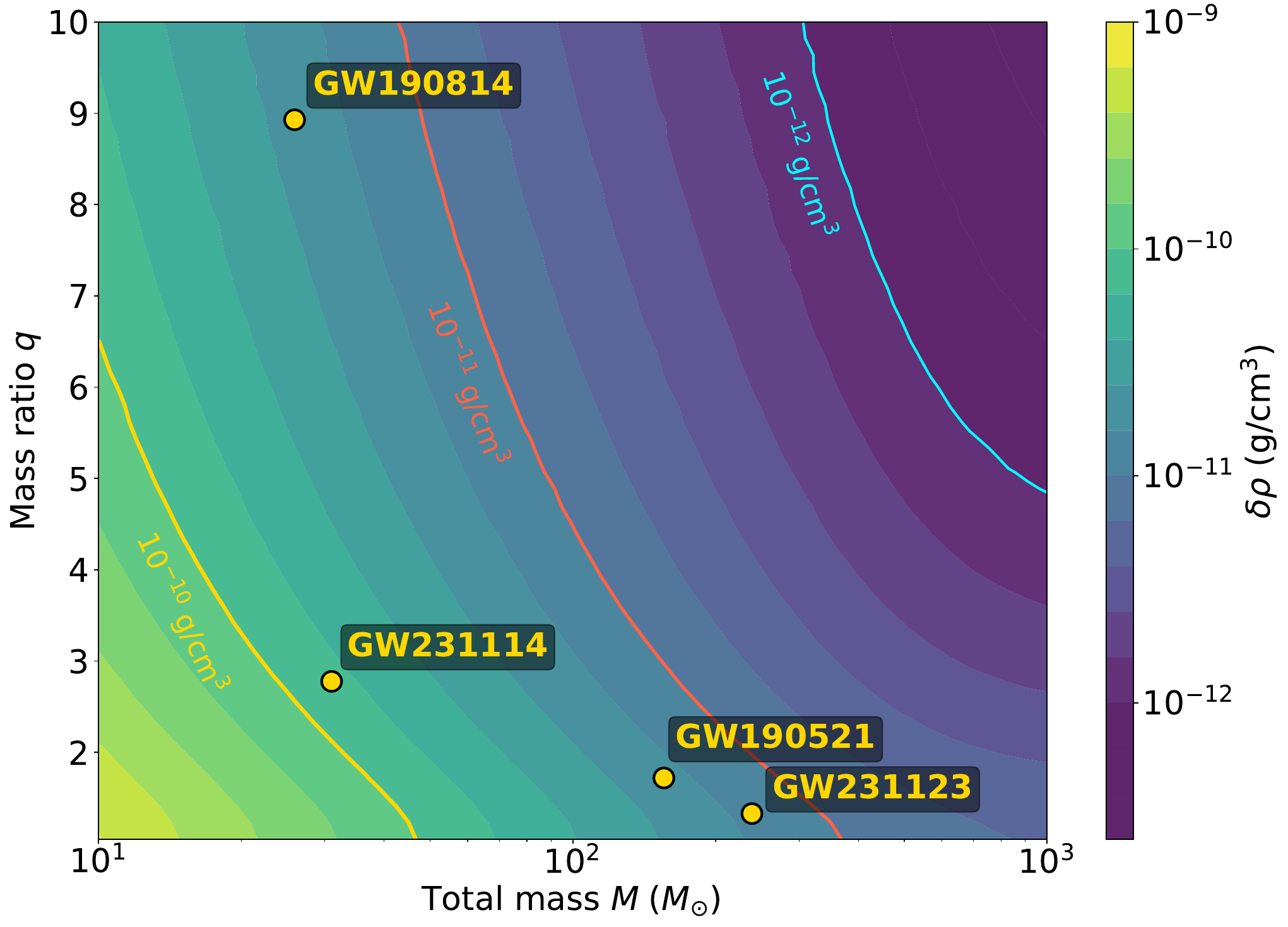}
    \caption{ The estimate precision $\delta\rho$ across \ac{bbh} systems with different total mass $M$ and mass ratio $q$ with the 3-month observation (solid lines). $\delta\rho$ equal to $10^{-12}/10^{-11}/10^{-10}\text{g/cm}^3$ are shown with blue/red/green lines, respectively. 
    The yellow dot represents the estimated precision of the gas density considering a GW190521-like event in 3 months.
    }
    \label{fig:FIM_M_q}
\end{figure}
%%==================================================
% Discuss the potential development of our method with different types of sBBH systems 

We further study how the precision of our gas density inference varies across different mass parameters of \ac{bbh} systems. 
As shown in Fig.~\ref{fig:FIM_M_q}, heavier and more asymmetric \acp{bbh} (higher total mass $M$, higher $q$) impose significantly tighter constraints on the ambient environment density. 
The enhanced precision enables the probing of a wider radial range in accretion disks: even for systems located farther from the disk center $>10^3r_s$, where the actual gas density variation between each 3-month segment become mild $(\Delta\rho \leq 10^{-10}\rm g/cm^3)$ — the high measurement accuracy $\delta\rho\sim10^{-12}\rm g/cm^3 $ remains sufficient to resolve such subtle variations $(\delta\rho< \Delta \rho)$. The recent detection of high-mass systems in GWTC-4, such as GW231123~\cite{LIGOScientific:2025rsn_GW231123}, which contains individual black holes exceeding $100M_{\odot}$, has better potential in probing \ac{agn} disk profile, as such systems reside squarely in the high-sensitivity region of the parameter space. 
Besides, we plot two \ac{gw} sources (GW190814\cite{LIGOScientific:2020zkf},GW231114\cite{LIGOScientific:2025slb}) which have relative larger mass ratio $q$.
%Thus, since the future O4 observations by \ac{lvk} could find more heavy and asymmetric \ac{bbh} systems. promising pathway to probing density profiles across a broad radial extent of \ac{agn} accretion disks.

\textit{Conclusion} - 
In this work, we have demonstrated for the first time that sequential \ac{gw} observations by space-borne detectors can be used to reconstruct the density profiles of \ac{agn} disks. 
By achieving unprecedented resolution down to $\delta\rho\sim 10^{-11}\rm g/cm^3$  via Bayesian inference (Fig.~\ref{fig:3D_posterior}), and by leveraging TianQin’s long-term observing capability, our method can trace systematic changes in the ambient density surrounding \acp{bbh}, ultimately constraining the global disk profile with a relative accuracy on the order of $2 \times 10^{-11}~\rm g/cm^3$.
Furthermore, we have shown that this sensitivity is enhanced for heavier and more asymmetric binary systems — as demonstrated by the detection of high-mass systems such as GW231123~\cite{LIGOScientific:2025rsn_GW231123}, along with high mass-ratio sources like GW190814~\cite{LIGOScientific:2020zkf} and GW231114~\cite{LIGOScientific:2025slb}, which exemplify the practical applicability of our approach.

In this letter,  we present an approach that is currently applicable to the AGN disk model with a fixed power-law density distribution (\(\rho \propto r^{-\alpha}\)). 
Beyond \ac{df} effect, we acknowledge potential contributions from other environmental effects (gas accretion, acceleration of central black holes), which temporarily lie below detection thresholds of space-borne \ac{gw} missions. 
Moreover, our method will be well-suited for probing ambient environment properties with next-generation space-borne gravitational-wave detectors offering higher sensitivity, such as DECIGO~\cite{DECIGO_2011}.

This methodology will expand multi-messenger observation coordination with TianQin, LISA, and other \ac{em} observatories. 
It will help reveal the deep connections between supermassive black hole accretion and galaxy evolution.

%%==================================

\begin{acknowledgments}
    Xiangyu thanks Enrico Barausse for his great help with learning the corresponding gravitational wave environmental effect knowledge and the program in this project. Jiandong Zhang and Jianwei Mei also contributed a meaningful discussion of our work.
    This work has been supported by the National Key
    Research and Development Program of China (No. 2023YFC2206700), 
    the Natural Science Foundation of China (Grant No. 12173104),
    the Natural Science Foundation of Guangdong Province of China (Grant No. 2022A1515011862),
    and the Fundamental Research Funds for the Central Universities, Sun Yat-sen University. 
\end{acknowledgments}

\bibliography{refpaper} %% bibfile 是自己的 bib 文件

@article{Lyu:2023ctt,
    author = "Lyu, Xiangyu and Li, En-Kun and Hu, Yi-Ming",
    title = "{Parameter estimation of stellar mass binary black holes in the network of TianQin and LISA}",
    primaryClass = "gr-qc",
    doi = "10.1103/PhysRevD.108.083023",
    journal = "Phys. Rev. D",
    volume = "108",
    number = "8",
    pages = "083023",
    year = "2023"
}

@article{LIGOScientific:2020zkf,
    author = "Abbott, R. and others",
    collaboration = "LIGO Scientific, Virgo",
    title = "{GW190814: Gravitational Waves from the Coalescence of a 23 Solar Mass Black Hole with a 2.6 Solar Mass Compact Object}",
    archivePrefix = "arXiv",
    primaryClass = "astro-ph.HE",
    reportNumber = "LIGO-P190814",
    doi = "10.3847/2041-8213/ab960f",
    journal = "Astrophys. J. Lett.",
    volume = "896",
    number = "2",
    pages = "L44",
    year = "2020"
}

@article{Enrico_accretion_inference_EMRI,
  title = {Influence of the hydrodynamic drag from an accretion torus on extreme mass-ratio inspirals},
  author = {Barausse, Enrico and Rezzolla, Luciano},
  journal = {Phys. Rev. D},
  volume = {77},
  issue = {10},
  pages = {104027},
  numpages = {22},
  year = {2008},
  month = {May},
  publisher = {American Physical Society},
  doi = {10.1103/PhysRevD.77.104027}
}

@article{FedrowPhysRevLett,
  title = {Gravitational Waves from Binary Black Hole Mergers inside Stars},
  author = {Fedrow, Joseph M. and Ott, Christian D. and Sperhake, Ulrich and Blackman, Jonathan and Haas, Roland and Reisswig, Christian and De Felice, Antonio},
  journal = {Phys. Rev. Lett.},
  volume = {119},
  issue = {17},
  pages = {171103},
  numpages = {5},
  year = {2017},
  month = {Oct},
  publisher = {American Physical Society},
  doi = {10.1103/PhysRevLett.119.171103}
}

@article{CanevaSantoro:2023aol,
    author = "Caneva Santoro, Giada and Roy, Soumen and Vicente, Rodrigo and Haney, Maria and Piccinni, Ornella Juliana and Del Pozzo, Walter and Martinez, Mario",
    title = "{First Constraints on Compact Binary Environments from LIGO-Virgo Data}",
    archivePrefix = "arXiv",
    primaryClass = "gr-qc",
    reportNumber = "LIGO DCC P2300301",
    doi = "10.1103/PhysRevLett.132.251401",
    journal = "Phys. Rev. Lett.",
    volume = "132",
    number = "25",
    pages = "251401",
    year = "2024"
}

@article{Liu:2021yoy,
    author = "Liu, Shuai and Zhu, Liang-Gui and Hu, Yi-Ming and Zhang, Jian-dong and Ji, Mu-Jie",
    title = "{Capability for detection of GW190521-like binary black holes with TianQin}",
    doi = "10.1103/PhysRevD.105.023019",
    journal = "Phys. Rev. D",
    volume = "105",
    number = "2",
    pages = "023019",
    year = "2022"
}

@ARTICLE{EHT2019ApJ,
       author = {{Event Horizon Telescope Collaboration} and {Akiyama}, Kazunori and {Alberdi}, Antxon and {Alef}, Walter and {Asada}, Keiichi and {Azulay}, Rebecca and {Baczko}, Anne-Kathrin and {Ball}, David and {Balokovi{\'c}}, Mislav and {Barrett}, John and {Bintley}, Dan and {Blackburn}, Lindy and {Boland}, Wilfred and {Bouman}, Katherine L. and {Bower}, Geoffrey C. and {Bremer}, Michael and {Brinkerink}, Christiaan D. and {Brissenden}, Roger and {Britzen}, Silke and {Broderick}, Avery E. and {Broguiere}, Dominique and {Bronzwaer}, Thomas and {Byun}, Do-Young and {Carlstrom}, John E. and {Chael}, Andrew and {Chan}, Chi-kwan and {Chatterjee}, Shami and {Chatterjee}, Koushik and {Chen}, Ming-Tang and {Chen}, Yongjun and {Cho}, Ilje and {Christian}, Pierre and {Conway}, John E. and {Cordes}, James M. and {Crew}, Geoffrey B. and {Cui}, Yuzhu and {Davelaar}, Jordy and {De Laurentis}, Mariafelicia and {Deane}, Roger and {Dempsey}, Jessica and {Desvignes}, Gregory and {Dexter}, Jason and {Doeleman}, Sheperd S. and {Eatough}, Ralph P. and {Falcke}, Heino and {Fish}, Vincent L. and {Fomalont}, Ed and {Fraga-Encinas}, Raquel and {Freeman}, William T. and {Friberg}, Per and {Fromm}, Christian M. and {G{\'o}mez}, Jos{\'e} L. and {Galison}, Peter and {Gammie}, Charles F. and {Garc{\'\i}a}, Roberto and {Gentaz}, Olivier and {Georgiev}, Boris and {Goddi}, Ciriaco and {Gold}, Roman and {Gu}, Minfeng and {Gurwell}, Mark and {Hada}, Kazuhiro and {Hecht}, Michael H. and {Hesper}, Ronald and {Ho}, Luis C. and {Ho}, Paul and {Honma}, Mareki and {Huang}, Chih-Wei L. and {Huang}, Lei and {Hughes}, David H. and {Ikeda}, Shiro and {Inoue}, Makoto and {Issaoun}, Sara and {James}, David J. and {Jannuzi}, Buell T. and {Janssen}, Michael and {Jeter}, Britton and {Jiang}, Wu and {Johnson}, Michael D. and {Jorstad}, Svetlana and {Jung}, Taehyun and {Karami}, Mansour and {Karuppusamy}, Ramesh and {Kawashima}, Tomohisa and {Keating}, Garrett K. and {Kettenis}, Mark and {Kim}, Jae-Young and {Kim}, Junhan and {Kim}, Jongsoo and {Kino}, Motoki and {Koay}, Jun Yi and {Koch}, Patrick M. and {Koyama}, Shoko and {Kramer}, Michael and {Kramer}, Carsten and {Krichbaum}, Thomas P. and {Kuo}, Cheng-Yu and {Lauer}, Tod R. and {Lee}, Sang-Sung and {Li}, Yan-Rong and {Li}, Zhiyuan and {Lindqvist}, Michael and {Liu}, Kuo and {Liuzzo}, Elisabetta and {Lo}, Wen-Ping and {Lobanov}, Andrei P. and {Loinard}, Laurent and {Lonsdale}, Colin and {Lu}, Ru-Sen and {MacDonald}, Nicholas R. and {Mao}, Jirong and {Markoff}, Sera and {Marrone}, Daniel P. and {Marscher}, Alan P. and {Mart{\'\i}-Vidal}, Iv{\'a}n and {Matsushita}, Satoki and {Matthews}, Lynn D. and {Medeiros}, Lia and {Menten}, Karl M. and {Mizuno}, Yosuke and {Mizuno}, Izumi and {Moran}, James M. and {Moriyama}, Kotaro and {Moscibrodzka}, Monika and {M{\"u}ller}, Cornelia and {Nagai}, Hiroshi and {Nagar}, Neil M. and {Nakamura}, Masanori and {Narayan}, Ramesh and {Narayanan}, Gopal and {Natarajan}, Iniyan and {Neri}, Roberto and {Ni}, Chunchong and {Noutsos}, Aristeidis and {Okino}, Hiroki and {Olivares}, H{\'e}ctor and {Ortiz-Le{\'o}n}, Gisela N. and {Oyama}, Tomoaki and {{\"O}zel}, Feryal and {Palumbo}, Daniel C.~M. and {Patel}, Nimesh and {Pen}, Ue-Li and {Pesce}, Dominic W. and {Pi{\'e}tu}, Vincent and {Plambeck}, Richard and {PopStefanija}, Aleksandar and {Porth}, Oliver and {Prather}, Ben and {Preciado-L{\'o}pez}, Jorge A. and {Psaltis}, Dimitrios and {Pu}, Hung-Yi and {Ramakrishnan}, Venkatessh and {Rao}, Ramprasad and {Rawlings}, Mark G. and {Raymond}, Alexander W. and {Rezzolla}, Luciano and {Ripperda}, Bart and {Roelofs}, Freek and {Rogers}, Alan and {Ros}, Eduardo and {Rose}, Mel and {Roshanineshat}, Arash and {Rottmann}, Helge and {Roy}, Alan L. and {Ruszczyk}, Chet and {Ryan}, Benjamin R. and {Rygl}, Kazi L.~J. and {S{\'a}nchez}, Salvador and {S{\'a}nchez-Arguelles}, David and {Sasada}, Mahito and {Savolainen}, Tuomas and {Schloerb}, F. Peter and {Schuster}, Karl-Friedrich and {Shao}, Lijing and {Shen}, Zhiqiang and {Small}, Des and {Sohn}, Bong Won and {SooHoo}, Jason and {Tazaki}, Fumie and {Tiede}, Paul and {Tilanus}, Remo P.~J. and {Titus}, Michael and {Toma}, Kenji and {Torne}, Pablo and {Trent}, Tyler and {Trippe}, Sascha and {Tsuda}, Shuichiro and {van Bemmel}, Ilse and {van Langevelde}, Huib Jan and {van Rossum}, Daniel R. and {Wagner}, Jan and {Wardle}, John and {Weintroub}, Jonathan and {Wex}, Norbert and {Wharton}, Robert and {Wielgus}, Maciek and {Wong}, George N. and {Wu}, Qingwen and {Young}, Ken and {Young}, Andr{\'e}},
        title = "{First M87 Event Horizon Telescope Results. I. The Shadow of the Supermassive Black Hole}",
      journal = "Astrophys. J. Lett.",
         year = 2019,
        month = apr,
       volume = {875},
       number = {1},
          eid = {L1},
        pages = {L1},
          doi = {10.3847/2041-8213/ab0ec7},
       adsurl = {https://ui.adsabs.harvard.edu/abs/2019ApJ...875L...1E}
}

@ARTICLE{EHT2022ApJ,
       author = {{Event Horizon Telescope Collaboration} and {Akiyama}, Kazunori and {Alberdi}, Antxon and {Alef}, Walter and {Algaba}, Juan Carlos and {Anantua}, Richard and {Asada}, Keiichi and {Azulay}, Rebecca and {Bach}, Uwe and {Baczko}, Anne-Kathrin and {Ball}, David and {Balokovi{\'c}}, Mislav and {Barrett}, John and {Baub{\"o}ck}, Michi and {Benson}, Bradford A. and {Bintley}, Dan and {Blackburn}, Lindy and {Blundell}, Raymond and {Bouman}, Katherine L. and {Bower}, Geoffrey C. and {Boyce}, Hope and {Bremer}, Michael and {Brinkerink}, Christiaan D. and {Brissenden}, Roger and {Britzen}, Silke and {Broderick}, Avery E. and {Broguiere}, Dominique and {Bronzwaer}, Thomas and {Bustamante}, Sandra and {Byun}, Do-Young and {Carlstrom}, John E. and {Ceccobello}, Chiara and {Chael}, Andrew and {Chan}, Chi-kwan and {Chatterjee}, Koushik and {Chatterjee}, Shami and {Chen}, Ming-Tang and {Chen}, Yongjun and {Cheng}, Xiaopeng and {Cho}, Ilje and {Christian}, Pierre and {Conroy}, Nicholas S. and {Conway}, John E. and {Cordes}, James M. and {Crawford}, Thomas M. and {Crew}, Geoffrey B. and {Cruz-Osorio}, Alejandro and {Cui}, Yuzhu and {Davelaar}, Jordy and {De Laurentis}, Mariafelicia and {Deane}, Roger and {Dempsey}, Jessica and {Desvignes}, Gregory and {Dexter}, Jason and {Dhruv}, Vedant and {Doeleman}, Sheperd S. and {Dougal}, Sean and {Dzib}, Sergio A. and {Eatough}, Ralph P. and {Emami}, Razieh and {Falcke}, Heino and {Farah}, Joseph and {Fish}, Vincent L. and {Fomalont}, Ed and {Ford}, H. Alyson and {Fraga-Encinas}, Raquel and {Freeman}, William T. and {Friberg}, Per and {Fromm}, Christian M. and {Fuentes}, Antonio and {Galison}, Peter and {Gammie}, Charles F. and {Garc{\'\i}a}, Roberto and {Gentaz}, Olivier and {Georgiev}, Boris and {Goddi}, Ciriaco and {Gold}, Roman and {G{\'o}mez-Ruiz}, Arturo I. and {G{\'o}mez}, Jos{\'e} L. and {Gu}, Minfeng and {Gurwell}, Mark and {Hada}, Kazuhiro and {Haggard}, Daryl and {Haworth}, Kari and {Hecht}, Michael H. and {Hesper}, Ronald and {Heumann}, Dirk and {Ho}, Luis C. and {Ho}, Paul and {Honma}, Mareki and {Huang}, Chih-Wei L. and {Huang}, Lei and {Hughes}, David H. and {Ikeda}, Shiro and {Impellizzeri}, C.~M. Violette and {Inoue}, Makoto and {Issaoun}, Sara and {James}, David J. and {Jannuzi}, Buell T. and {Janssen}, Michael and {Jeter}, Britton and {Jiang}, Wu and {Jim{\'e}nez-Rosales}, Alejandra and {Johnson}, Michael D. and {Jorstad}, Svetlana and {Joshi}, Abhishek V. and {Jung}, Taehyun and {Karami}, Mansour and {Karuppusamy}, Ramesh and {Kawashima}, Tomohisa and {Keating}, Garrett K. and {Kettenis}, Mark and {Kim}, Dong-Jin and {Kim}, Jae-Young and {Kim}, Jongsoo and {Kim}, Junhan and {Kino}, Motoki and {Koay}, Jun Yi and {Kocherlakota}, Prashant and {Kofuji}, Yutaro and {Koch}, Patrick M. and {Koyama}, Shoko and {Kramer}, Carsten and {Kramer}, Michael and {Krichbaum}, Thomas P. and {Kuo}, Cheng-Yu and {La Bella}, Noemi and {Lauer}, Tod R. and {Lee}, Daeyoung and {Lee}, Sang-Sung and {Leung}, Po Kin and {Levis}, Aviad and {Li}, Zhiyuan and {Lico}, Rocco and {Lindahl}, Greg and {Lindqvist}, Michael and {Lisakov}, Mikhail and {Liu}, Jun and {Liu}, Kuo and {Liuzzo}, Elisabetta and {Lo}, Wen-Ping and {Lobanov}, Andrei P. and {Loinard}, Laurent and {Lonsdale}, Colin J. and {Lu}, Ru-Sen and {Mao}, Jirong and {Marchili}, Nicola and {Markoff}, Sera and {Marrone}, Daniel P. and {Marscher}, Alan P. and {Mart{\'\i}-Vidal}, Iv{\'a}n and {Matsushita}, Satoki and {Matthews}, Lynn D. and {Medeiros}, Lia and {Menten}, Karl M. and {Michalik}, Daniel and {Mizuno}, Izumi and {Mizuno}, Yosuke and {Moran}, James M. and {Moriyama}, Kotaro and {Moscibrodzka}, Monika and {M{\"u}ller}, Cornelia and {Mus}, Alejandro and {Musoke}, Gibwa and {Myserlis}, Ioannis and {Nadolski}, Andrew and {Nagai}, Hiroshi and {Nagar}, Neil M. and {Nakamura}, Masanori and {Narayan}, Ramesh and {Narayanan}, Gopal and {Natarajan}, Iniyan and {Nathanail}, Antonios and {Fuentes}, Santiago Navarro and {Neilsen}, Joey and {Neri}, Roberto and {Ni}, Chunchong and {Noutsos}, Aristeidis and {Nowak}, Michael A. and {Oh}, Junghwan and {Okino}, Hiroki and {Olivares}, H{\'e}ctor and {Ortiz-Le{\'o}n}, Gisela N. and {Oyama}, Tomoaki and {{\"O}zel}, Feryal and {Palumbo}, Daniel C.~M. and {Paraschos}, Georgios Filippos and {Park}, Jongho and {Parsons}, Harriet and {Patel}, Nimesh and {Pen}, Ue-Li and {Pesce}, Dominic W. and {Pi{\'e}tu}, Vincent and {Plambeck}, Richard and {PopStefanija}, Aleksandar and {Porth}, Oliver and {P{\"o}tzl}, Felix M. and {Prather}, Ben and {Preciado-L{\'o}pez}, Jorge A. and {Psaltis}, Dimitrios},
        title = "{First Sagittarius A* Event Horizon Telescope Results. I. The Shadow of the Supermassive Black Hole in the Center of the Milky Way}",
      journal = "Astrophys. J. Lett.",
         year = 2022,
        month = may,
       volume = {930},
       number = {2},
          eid = {L12},
        pages = {L12},
          doi = {10.3847/2041-8213/ac6674},
       adsurl = {https://ui.adsabs.harvard.edu/abs/2022ApJ...930L..12E}
}

@article{foreman-mackey_emcee_2013,
    author = "Foreman-Mackey, Daniel and Hogg, David W. and Lang, Dustin and Goodman, Jonathan",
    title = "{emcee: The MCMC Hammer}",
    eprint = "1202.3665",
    archivePrefix = "arXiv",
    primaryClass = "astro-ph.IM",
    doi = "10.1086/670067",
    journal = "Publ. Astron. Soc. Pac.",
    volume = "125",
    pages = "306--312",
    year = "2013"
}

@article{LIGOScientific:2021usb,
    author = "Abbott, R. and others",
    collaboration = "LIGO Scientific, VIRGO",
    title = "{GWTC-2.1: Deep extended catalog of compact binary coalescences observed by LIGO and Virgo during the first half of the third observing run}",
    archivePrefix = "arXiv",
    primaryClass = "gr-qc",
    reportNumber = "LIGO-P2100063",
    doi = "10.1103/PhysRevD.109.022001",
    journal = "Phys. Rev. D",
    volume = "109",
    number = "2",
    pages = "022001",
    year = "2024"
}

@article{liu_science_2020,
    author = "Liu, Shuai and Hu, Yi-Ming and Zhang, Jian-dong and Mei, Jianwei",
    title = "{Science with the TianQin observatory: Preliminary results on stellar-mass binary black holes}",
    archivePrefix = "arXiv",
    primaryClass = "astro-ph.HE",
    doi = "10.1103/PhysRevD.101.103027",
    journal = "Phys. Rev. D",
    volume = "101",
    number = "10",
    pages = "103027",
    year = "2020"
}

@article{luo_tianqin:_2016,
    author = "Luo, Jun and others",
    collaboration = "TianQin",
    title = "{TianQin: a space-borne gravitational wave detector}",

    primaryClass = "astro-ph.IM",
    doi = "10.1088/0264-9381/33/3/035010",
    journal = "Class. Quant. Grav.",
    volume = "33",
    number = "3",
    pages = "035010",
    year = "2016"
}

@article{marsat_exploring_2021,
    author = "Marsat, Sylvain and Baker, John G. and Dal Canton, Tito",
    title = "{Exploring the Bayesian parameter estimation of binary black holes with LISA}",
    archivePrefix = "arXiv",
    primaryClass = "gr-qc",
    doi = "10.1103/PhysRevD.103.083011",
    journal = "Phys. Rev. D",
    volume = "103",
    number = "8",
    pages = "083011",
    year = "2021"
}

@article{LIGOScientific:2025rsn_GW231123,
    author = "Abac, A. G. and others",
    collaboration = "LIGO Scientific, VIRGO, KAGRA",
    title = "{GW231123: a Binary Black Hole Merger with Total Mass 190-265 $M_{\odot}$}",
    eprint = "2507.08219",
    archivePrefix = "arXiv",
    primaryClass = "astro-ph.HE",
    reportNumber = "DCC: P2500026-v6",
    month = "7",
    year = "2025"
}

@article{Tamanini:2016zlh,
    author = "Tamanini, Nicola and Caprini, Chiara and Barausse, Enrico and Sesana, Alberto and Klein, Antoine and Petiteau, Antoine",
    title = "{Science with the space-based interferometer eLISA. III: Probing the expansion of the Universe using gravitational wave standard sirens}",

    archivePrefix = "arXiv",
    primaryClass = "astro-ph.CO",
    doi = "10.1088/1475-7516/2016/04/002",
    journal = "JCAP",
    volume = "04",
    pages = "002",
    year = "2016"
}

@article{IMRPhenomPv2,
  title = {Frequency-domain gravitational waves from nonprecessing black-hole binaries. II. A phenomenological model for the advanced detector era},
  author = {Khan, Sebastian and Husa, Sascha and Hannam, Mark and Ohme, Frank and P\"urrer, Michael and Forteza, Xisco Jim\'enez and Boh\'e, Alejandro},
  journal = {Phys. Rev. D},
  volume = {93},
  issue = {4},
  pages = {044007},
  numpages = {27},
  year = {2016},
  month = {Feb},
  publisher = {American Physical Society},
  doi = {10.1103/PhysRevD.93.044007}
}

@article{DuttaRoy:2025gnu,
    author = "Dutta Roy, Poulami and Mahapatra, Parthapratim and Samajdar, Anuradha and Arun, K. G.",
    title = "{Identifying intermediate mass binary black hole mergers in AGN disks using LISA}",

    archivePrefix = "arXiv",
    primaryClass = "astro-ph.HE",
    doi = "10.1103/PhysRevD.111.104047",
    journal = "Phys. Rev. D",
    volume = "111",
    number = "10",
    pages = "104047",
    year = "2025"
}

@article{LIGOScientific:2020ufj,
    author = "Abbott, R. and others",
    collaboration = "LIGO Scientific, Virgo",
    title = "{Properties and Astrophysical Implications of the 150 M$_\odot$ Binary Black Hole Merger GW190521}",
    archivePrefix = "arXiv",
    primaryClass = "astro-ph.HE",
    reportNumber = "LIGO-P2000021",
    doi = "10.3847/2041-8213/aba493",
    journal = "Astrophys. J. Lett.",
    volume = "900",
    number = "1",
    pages = "L13",
    year = "2020"
}

@article{LIGOScientific:2020iuh,
    author = "Abbott, R. and others",
    collaboration = "LIGO Scientific, Virgo",
    title = "{GW190521: A Binary Black Hole Merger with a Total Mass of $150  M_{\odot}$}",
    archivePrefix = "arXiv",
    primaryClass = "gr-qc",
    doi = "10.1103/PhysRevLett.125.101102",
    journal = "Phys. Rev. Lett.",
    volume = "125",
    number = "10",
    pages = "101102",
    year = "2020"
}

@article{Yang:2024tje,
    author = "Yang, Shu-Cheng and Han, Wen-Biao and Tagawa, Hiromichi and Li, Song and Jiang, Ye and Shen, Ping and Yun, Qianyun and Zhang, Chen and Zhong, Xing-Yu",
    title = "{Indication for a Compact Object Next to a LIGO{\textendash}Virgo Binary Black Hole Merger}",
    eprint = "2401.01743",
    archivePrefix = "arXiv",
    primaryClass = "astro-ph.HE",
    doi = "10.3847/2041-8213/adeaad",
    journal = "Astrophys. J. Lett.",
    volume = "988",
    number = "2",
    pages = "L41",
    year = "2025"
}

@article{cutler_gravitational_1994,
    author = "Cutler, Curt and Flanagan, Eanna E.",
    title = "{Gravitational waves from merging compact binaries: How accurately can one extract the binary's parameters from the inspiral wave form?}",
    archivePrefix = "arXiv",
    reportNumber = "GRP-369",
    doi = "10.1103/PhysRevD.49.2658",
    journal = "Phys. Rev. D",
    volume = "49",
    pages = "2658--2697",
    year = "1994"
}

@article{sesana_prospects_2016,
    author = "Sesana, Alberto",
    title = "{Prospects for Multiband Gravitational-Wave Astronomy after GW150914}",
    archivePrefix = "arXiv",
    primaryClass = "gr-qc",
    doi = "10.1103/PhysRevLett.116.231102",
    journal = "Phys. Rev. Lett.",
    volume = "116",
    number = "23",
    pages = "231102",
    year = "2016"
}

@article{Zhu:2021bpp,
    author = "Zhu, Liang-Gui and Xie, Ling-Hua and Hu, Yi-Ming and Liu, Shuai and Li, En-Kun and Napolitano, Nicola R. and Tang, Bai-Tian and Zhang, Jian-dong and Mei, Jianwei",
    title = "{Constraining the Hubble constant to a precision of about 1\% using multi-band dark standard siren detections}",
    archivePrefix = "arXiv",
    primaryClass = "astro-ph.CO",
    doi = "10.1007/s11433-021-1859-9",
    journal = "Sci. China Phys. Mech. Astron.",
    volume = "65",
    number = "5",
    pages = "259811",
    year = "2022"
}

@article{Toubiana:2022vpp,
    author = "Toubiana, Alexandre and Babak, Stanislav and Marsat, Sylvain and Ossokine, Sergei",
    title = "{Detectability and parameter estimation of GWTC-3 events with LISA}",
    archivePrefix = "arXiv",
    primaryClass = "gr-qc",
    doi = "10.1103/PhysRevD.106.104034",
    journal = "Phys. Rev. D",
    volume = "106",
    number = "10",
    pages = "104034",
    year = "2022"
}

@article{buscicchio_bayesian_2021,
    author = "Buscicchio, Riccardo and Klein, Antoine and Roebber, Elinore and Moore, Christopher J. and Gerosa, Davide and Finch, Eliot and Vecchio, Alberto",
    title = "{Bayesian parameter estimation of stellar-mass black-hole binaries with LISA}",

    archivePrefix = "arXiv",
    primaryClass = "astro-ph.HE",
    doi = "10.1103/PhysRevD.104.044065",
    journal = "Phys. Rev. D",
    volume = "104",
    number = "4",
    pages = "044065",
    year = "2021"
}

@article{Miralda-Escude:2000kqv,
    author = "Miralda-Escude, Jordi and Gould, Andrew",
    title = "{A cluster of black holes at the galactic center}",

    archivePrefix = "arXiv",
    doi = "10.1086/317837",
    journal = "Astrophys. J.",
    volume = "545",
    pages = "847",
    year = "2000"
}

@article{Inayoshi:2017hgw,
    author = "Inayoshi, Kohei and Tamanini, Nicola and Caprini, Chiara and Haiman, Zolt\'an",
    title = "{Probing stellar binary black hole formation in galactic nuclei via the imprint of their center of mass acceleration on their gravitational wave signal}",
    archivePrefix = "arXiv",
    primaryClass = "astro-ph.HE",
    doi = "10.1103/PhysRevD.96.063014",
    journal = "Phys. Rev. D",
    volume = "96",
    number = "6",
    pages = "063014",
    year = "2017"
}

@article{Sesana:2017vsj,
    author = "Sesana, Alberto",
    editor = "Giardini, Domencio and Jetzer, Philippe",
    title = "{Multi-band gravitational wave astronomy: science with joint space- and ground-based observations of black hole binaries}",

    archivePrefix = "arXiv",
    primaryClass = "astro-ph.HE",
    doi = "10.1088/1742-6596/840/1/012018",
    journal = "J. Phys. Conf. Ser.",
    volume = "840",
    number = "1",
    pages = "012018",
    year = "2017"
}

@article{Toubiana:2020vtf,
    author = "Toubiana, Alexandre and Marsat, Sylvian and Barausse, Enrico and Babak, Stanislav and Baker, John",
    title = "{Tests of general relativity with stellar-mass black hole binaries observed by LISA}",
    eprint = "2004.03626",
    archivePrefix = "arXiv",
    primaryClass = "gr-qc",
    doi = "10.1103/PhysRevD.101.104038",
    journal = "Phys. Rev. D",
    volume = "101",
    number = "10",
    pages = "104038",
    year = "2020"
}

@article{Fabian:2012xr,
    author = "Fabian, A. C.",
    title = "{Observational Evidence of AGN Feedback}",
    archivePrefix = "arXiv",
    primaryClass = "astro-ph.CO",
    doi = "10.1146/annurev-astro-081811-125521",
    journal = "Ann. Rev. Astron. Astrophys.",
    volume = "50",
    pages = "455--489",
    year = "2012"
}

@ARTICLE{1978Natur_Shields,
       author = {{Shields}, G.~A.},
        title = "{Thermal continuum from accretion disks in quasars}",
      journal = {\nat},
         year = 1978,
        month = apr,
       volume = {272},
       number = {5655},
        pages = {706-708},
          doi = {10.1038/272706a0}
}

@ARTICLE{2023NatAs_Tang,
       author = {{Tang}, Ji-Jia and {Wolf}, Christian and {Tonry}, John},
        title = "{Universality in the random walk structure function of luminous quasi-stellar objects}",
      journal = {Nat. Astron.},
         year = 2023,
        month = apr,
       volume = {7},
        pages = {473-480},
          doi = {10.1038/s41550-022-01885-8}
}

@ARTICLE{2021Sci_Burke,
       author = {{Burke}, Colin J. and {Shen}, Yue and {Blaes}, Omer and {Gammie}, Charles F. and {Horne}, Keith and {Jiang}, Yan-Fei and {Liu}, Xin and {McHardy}, Ian M. and {Morgan}, Christopher W. and {Scaringi}, Simone and {Yang}, Qian},
        title = "{A characteristic optical variability time scale in astrophysical accretion disks}",
      journal = {Science},
         year = 2021,
        month = aug,
       volume = {373},
       number = {6556},
        pages = {789-792},
          doi = {10.1126/science.abg9933},
archivePrefix = {arXiv},
       eprint = {2108.05389},
 primaryClass = {astro-ph.GA}
}

@ARTICLE{2009ApJ,
       author = {{Kelly}, Brandon C. and {Bechtold}, Jill and {Siemiginowska}, Aneta},
        title = "{Are the Variations in Quasar Optical Flux Driven by Thermal Fluctuations?}",
      journal = {\apj},
         year = 2009,
        month = jun,
       volume = {698},
       number = {1},
        pages = {895-910},
          doi = {10.1088/0004-637X/698/1/895},
archivePrefix = {arXiv},
       eprint = {0903.5315},
 primaryClass = {astro-ph.CO}
}

@article{Kormendy:2013dxa,
    author = "Kormendy, John and Ho, Luis C.",
    title = "{Coevolution (Or Not) of Supermassive Black Holes and Host Galaxies}",
    eprint = "1304.7762",
    archivePrefix = "arXiv",
    primaryClass = "astro-ph.CO",
    doi = "10.1146/annurev-astro-082708-101811",
    journal = "Ann. Rev. Astron. Astrophys.",
    volume = "51",
    pages = "511--653",
    year = "2013"
}

@article{DECIGO_2011,
author = "Kawamura, Seiji and others",
doi = {10.1088/0264-9381/28/9/094011},
year = {2011},
month = {apr},
publisher = {},
volume = {28},
number = {9},
pages = {094011},
title = {The Japanese space gravitational wave antenna: DECIGO},
journal = {Class. Quantum Grav.},
}

@article{Rees:1984si,
    author = "Rees, Martin J.",
    title = "{Black Hole Models for Active Galactic Nuclei}",
    doi = "10.1146/annurev.aa.22.090184.002351",
    journal = "Ann. Rev. Astron. Astrophys.",
    volume = "22",
    pages = "471--506",
    year = "1984"
}

@article{baan_h2o_2022,
	title = {H2O {MegaMaser} emission in {NGC} 4258 indicative of a periodic disc instability},
	volume = {6},
	issn = {2397-3366},
	doi = {10.1038/s41550-022-01706-y},
	pages = {976--983},
	number = {8},
	journal = {Nat. Astron.},
	author = {Baan, Willem A. and An, Tao and Henkel, Christian and Imai, Hiroshi and Kostenko, Vladimir and Sobolev, Andrej},
	year= "2022"
}

@article{LIGOScientific:2025slb,
    author = "Abac, A. G. and others",
    collaboration = "LIGO Scientific, VIRGO, KAGRA",
    title = "{GWTC-4.0: Updating the Gravitational-Wave Transient Catalog with Observations from the First Part of the Fourth LIGO-Virgo-KAGRA Observing Run}",
    eprint = "2508.18082",
    archivePrefix = "arXiv",
    primaryClass = "gr-qc",
    reportNumber = "LIGO-P2400386",
    month = "8",
    year = "2025"
}

@article{Pounds:2003tm,
    author = "Pounds, K. A. and Reeves, J. N. and Page, Kim L. and Wynn, G. A. and O'Brien, P. T.",
    title = "{Fe K emission and absorption features in XMM-Newton spectra of Mkn 766: Evidence for reprocessing in flare ejecta}",
    archivePrefix = "arXiv",
    doi = "10.1046/j.1365-8711.2003.06611.x",
    journal = "Mon. Not. Roy. Astron. Soc.",
    volume = "342",
    pages = "1147",
    year = "2003"
}

@article{Jiang_2025,
doi = {10.3847/2041-8213/adc456},
year = {2025},
month = {apr},
publisher = {The American Astronomical Society},
volume = {983},
number = {1},
pages = {L18},
author = {Jiang, Ning and Pan, Zhen},
title = {Embers of Active Galactic Nuclei: Tidal Disruption Events and Quasiperiodic Eruptions},
journal = {Astrophys. J. Lett.}
}

@article{Linial:2023nqs,
    author = "Linial, Itai and Metzger, Brian D.",
    title = "{EMRI + TDE = QPE: Periodic X-Ray Flares from Star\textendash{}Disk Collisions in Galactic Nuclei}",
    eprint = "2303.16231",
    archivePrefix = "arXiv",
    primaryClass = "astro-ph.HE",
    doi = "10.3847/1538-4357/acf65b",
    journal = "Astrophys. J.",
    volume = "957",
    number = "1",
    pages = "34",
    year = "2023"
}

@ARTICLE{Morris1993ApJ,
       author = {{Morris}, Mark},
        title = "{Massive Star Formation near the Galactic Center and the Fate of the Stellar Remnants}",
      journal = {Astrophys. J.},
         year = 1993,
        month = may,
       volume = {408},
        pages = {496},
          doi = {10.1086/172607},
       adsurl = {https://ui.adsabs.harvard.edu/abs/1993ApJ...408..496M}
}

@article{hailey_density_2018,
	title = {A density cusp of quiescent X-ray binaries in the central parsec of the Galaxy},
	volume = {556},
	issn = {1476-4687},
	doi = {10.1038/nature25029},
	pages = {70--73},
    year = 2018 ,
	number = {7699},
	journal = {Nature},
	author = {Hailey, Charles J. and Mori, Kaya and Bauer, Franz E. and Berkowitz, Michael E. and Hong, Jaesub and Hord, Benjamin J.},
	date = {2018-04-01}
}

@article{Generozov:2018niv,
    author = "Generozov, A. and Stone, N. C. and Metzger, B. D. and Ostriker, J. P.",
    title = "{An overabundance of black hole X-ray binaries in the Galactic Centre from tidal captures}",

    archivePrefix = "arXiv",
    primaryClass = "astro-ph.HE",
    doi = "10.1093/mnras/sty1262",
    journal = "Mon. Not. Roy. Astron. Soc.",
    volume = "478",
    number = "3",
    pages = "4030--4051",
    year = "2018"
}

@article{Barausse_environmental_astrophysics,
  title = {Can environmental effects spoil precision gravitational-wave astrophysics?},
  author = {Barausse, Enrico and Cardoso, Vitor and Pani, Paolo},
  journal = {Phys. Rev. D},
  volume = {89},
  issue = {10},
  pages = {104059},
  numpages = {60},
  year = {2014},
  month = {May},
  publisher = {American Physical Society},
  doi = {10.1103/PhysRevD.89.104059}
}

@article{Toubiana:2020drf,
    author = "Toubiana, Alexandre and others",
    title = "{Detectable environmental effects in GW190521-like black-hole binaries with LISA}",
    archivePrefix = "arXiv",
    primaryClass = "astro-ph.HE",
    doi = "10.1103/PhysRevLett.126.101105",
    journal = "Phys. Rev. Lett.",
    volume = "126",
    number = "10",
    pages = "101105",
    year = "2021"
}

@ARTICLE{Toomre1964ApJ,
       author = {{Toomre}, A.},
        title = "{On the gravitational stability of a disk of stars.}",
      journal = "Astrophys. J.",
         year = 1964,
        month = may,
       volume = {139},
        pages = {1217-1238},
          doi = {10.1086/147861},
       adsurl = {https://ui.adsabs.harvard.edu/abs/1964ApJ...139.1217T}
}

@article{Farmer:2019jed,
author = {{Farmer}, R. and {Renzo}, M. and {de Mink}, S.~E. and {Marchant}, P. and {Justham}, S.},
        title = "{Mind the Gap: The Location of the Lower Edge of the Pair-instability Supernova Black Hole Mass Gap}",
      journal = {\apj},
         year = 2019,
        month = dec,
       volume = {887},
       number = {1},
          eid = {53},
        pages = {53},
          doi = {10.3847/1538-4357/ab518b},
archivePrefix = {arXiv},
       eprint = {1910.12874},
 primaryClass = {astro-ph.SR},
       adsurl = {https://ui.adsabs.harvard.edu/abs/2019ApJ...887...53F}
}

@article{Mapelli:2019ipt,
    author = "Mapelli, Michela and Spera, Mario and Montanari, Enrico and Limongi, Marco and Chieffi, Alessandro and Giacobbo, Nicola and Bressan, Alessandro and Bouffanais, Yann",
    title = "{Impact of the Rotation and Compactness of Progenitors on the Mass of Black Holes}",
    primaryClass = "astro-ph.HE",
    doi = "10.3847/1538-4357/ab584d",
    journal = "Astrophys. J.",
    volume = "888 ",
    pages = "76",
    year = "2020"
}

@article{Marchant:2020haw,
    author = "Marchant, Pablo and Moriya, Takashi",
    title = "{The impact of stellar rotation on the black hole mass-gap from pair-instability supernovae}",
    primaryClass = "astro-ph.HE",
    doi = "10.1051/0004-6361/202038902",
    journal = "Astron. Astrophys.",
    volume = "640",
    pages = "L18",
    year = "2020"
}

@article{Woosley:2021xba,
    author = "Woosley, S. E. and Heger, Alexander",
    title = "{The Pair-Instability Mass Gap for Black Holes}",
    doi = "10.3847/2041-8213/abf2c4",
    journal = "Astrophys. J. Lett.",
    volume = "912",
    number = "2",
    pages = "L31",
    year = "2021"
}

@article{Hendriks:2023yrw,
    author = "Hendriks, D. D. and van Son, L. A. C. and Renzo, M. and Izzard, R. G. and Farmer, R.",
    title = "{Pulsational pair-instability supernovae in gravitational-wave and electromagnetic transients}",
    doi = "10.1093/mnras/stad2857",
    journal = "Mon. Not. Roy. Astron. Soc.",
    volume = "526",
    number = "3",
    pages = "4130--4147",
    year = "2023"
}

@article{Thompson:2005mf,
    author = "Thompson, Todd A. and Quataert, Eliot and Murray, Norm",
    title = "{Radiation pressure supported starburst disks and AGN fueling}",
    eprint = "astro-ph/0503027",
    archivePrefix = "arXiv",
    doi = "10.1086/431923",
    journal = "Astrophys. J.",
    volume = "630",
    pages = "167--185",
    year = "2005"
}

@article{Pringle:1981ds,
    author = "Pringle, J. E.",
    title = "{Accretion discs in astrophysics}",
    doi = "10.1146/annurev.aa.19.090181.001033",
    journal = "Ann. Rev. Astron. Astrophys.",
    volume = "19",
    pages = "137--160",
    year = "1981"
}

@article{Sirko:2002ex,
    author = "Sirko, Edwin and Goodman, Jeremy",
    title = "{Spectral energy distributions of selfgravitating QSO discs}",
    archivePrefix = "arXiv",
    doi = "10.1046/j.1365-8711.2003.06431.x",
    journal = "Mon. Not. Roy. Astron. Soc.",
    volume = "341",
    pages = "501",
    year = "2003"
}

@article{Caputo:2020irr,
    author = "Caputo, Andrea and Sberna, Laura and Toubiana, Alexandre and Babak, Stanislav and Barausse, Enrico and Marsat, Sylvain and Pani, Paolo",
    title = "{Gravitational-wave detection and parameter estimation for accreting black-hole binaries and their electromagnetic counterpart}",
    archivePrefix = "arXiv",
    primaryClass = "astro-ph.HE",
    doi = "10.3847/1538-4357/ab7b66",
    journal = "Astrophys. J.",
    volume = "892",
    number = "2",
    pages = "90",
    year = "2020"
}

@article{Kocsis_accretion_DF,
  title = {Observable signatures of extreme mass-ratio inspiral black hole binaries embedded in thin accretion disks},
  author = {Kocsis, Bence and Yunes, Nicol\'as and Loeb, Abraham},
  journal = {Phys. Rev. D},
  volume = {84},
  issue = {2},
  pages = {024032},
  numpages = {41},
  year = {2011},
  month = {Jul},
  publisher = {American Physical Society},
  doi = {10.1103/PhysRevD.84.024032}
}

\appendix
% Introduce how to deal with density during the three months
\section{Assessment of the constant-density approximation bias }

We assess the systematic bias introduced by approximating the gas density as constant over each three-month observational segment, equal to its value at the segment’s midpoint. 
To quantify this bias, we develop a method to estimate the resulting uncertainty in the inferred radial distance $r_i(t)$ of the \acp{bbh} in the accretion disk.

First, we define an effective mean density $\bar{\rho}_{\rm seg}$ for each three-month segment such that the integrated inner product under this constant density equals that under the true time-varying density $\rho_i(t)$. 
Specifically, $\bar{\rho}_{\rm seg}$ satisfies:
\begin{equation}
    \int_{f_0}^{f_{\rm3mon}} \tilde{\phi}_{\rm DF}(f,\bar{\rho}_{\rm seg})df = \int_{f_0}^{f_{\rm 3mon}} \tilde{\phi}_{\rm DF}(f,\rho_i(t(f)))df,
\end{equation}
where the time-frequency relation $t(f) = -\frac{1}{2\pi} \frac{d\Phi}{df}$.

We then compute the time offset $\Delta t$ between the instant when $\rho(t) = \bar{\rho}$ and the nominal midpoint $t_0$. 
Using the phase-derived relation $t(f)$, we reconstruct the density variation over the full five-year observation. 
For each of the 20 segments, the offset $\Delta t_i$ is calculated. 
The distribution of these offsets is modelled as Gaussian, and its standard deviation $\sigma_{\Delta t}$ can be estimated.

Finally, we convert the temporal uncertainty $\sigma_{\Delta t}$ into a radial uncertainty $\delta r$ using the orbital kinematics of the binary. 
The final result reports the inferred radial location as $r_i \pm \delta r_i$, accounting for the systematic error introduced by the constant-density assumption.

% \section{The chosen sound speed values in the Bayesian inference process}

% In estimating TianQin's capability of constraining environmental effect, we assume the sound speed in the \ac{df} effect to be a constant $(c_s=0.02)$ during the three-month \ac{gw} observation.
% But $c_s \approx v_{\rm orb} (H/r)$~\cite{Frank_King_Raine_2002}(i.e. disk height $H$ at radius $r$) will vary with the movement of \ac{bbh} system in disk, as we assume $c_s$ is related with the galactocentric orbital velocity $v_{\rm orb}=\sqrt{M_{\rm SMBH}\left( \frac{2}{r} - \frac{1}{a} \right)}$, $a$ is the semimajor axis of orbit of barycenter of \ac{bbh}, and aspect ratio $H/r$, and following ~\citet{Graham:2020gwr}'s work, the disk's aspect ratio $H/r \sim 0.01$.
% However, we simplify this condition by assuming the $c_s^i=0.02c,i=1,2,\dots,20$ as we found the variation of the value of sound speed wouldn't significantly affect the fundamental observables, the phase shift $\tilde{\phi}_{\rm -5.5PN}^i, i=1,2,\dots,20$ caused by \ac{df} effect from sequential observations.

\end{document}